\documentclass[aps,twocolumn,prb,nopacs,floatfix,superscriptaddress]{revtex4}%
\pdfoutput=1 
\usepackage{graphicx}
\usepackage{amsmath}
\usepackage{amsfonts}
\usepackage{amssymb}
\usepackage{placeins}
\usepackage{helvet}
\usepackage{dcolumn}
\usepackage{nicefrac}

\newcommand{\GHS}{GaCu$_{3}$(OH)$_{6}$Cl$_{2}$}
\begin{document}
\title{Theoretical prediction of a strongly correlated Dirac metal}
\author{I. I. Mazin}
\affiliation{Code 6393, Naval Research Laboratory, Washington, DC 20375, USA}
\author{Harald O. Jeschke}
\affiliation{Institut f\"ur Theoretische Physik, Goethe-Universit\"at Frankfurt,
Max-von-Laue-Strasse 1, 60438 Frankfurt am Main, Germany}
\author{Frank Lechermann}
\affiliation{I. Institut f\"ur Theoretische Physik, Universit\"at Hamburg, D-20355 Hamburg, Germany}
\author{Hunpyo Lee}
\affiliation{Institut f\"ur Theoretische Physik, Goethe-Universit\"at Frankfurt,
Max-von-Laue-Strasse 1, 60438 Frankfurt am Main, Germany}
\author{Mario Fink}
\affiliation{Institut f\"ur Theoretische Physik I, Universit\"at W\"urzburg, am Hubland,
97074 W\"urzburg, Germany}
\author{Ronny Thomale}
\affiliation{Institut f\"ur Theoretische Physik I, Universit\"at W\"urzburg, am Hubland,
97074 W\"urzburg, Germany}
\author{Roser Valent\'\i}
\affiliation{Institut f\"ur Theoretische Physik, Goethe-Universit\"at Frankfurt,
Max-von-Laue-Strasse 1, 60438 Frankfurt am Main, Germany}
\date{\today}

\begin{abstract}
  \textbf{Recently, the most intensely studied objects in the
    electronic theory of solids have been strongly correlated systems
    and graphene. However, the fact that the Dirac bands in graphene
    are made up of $sp^{2}$-electrons, which are subject to neither
    strong Hubbard repulsion $U$ nor strong Hund's rule coupling $J$
    creates certain limitations in terms of novel, interaction-induced
    physics that could be derived from Dirac points. Here we propose
    {\GHS} (Ga-substituted herbertsmithite) as a
    correlated Dirac-Kagome metal combining Dirac electrons, strong
    interactions and frustrated magnetism. Using density functional
    theory (DFT), we calculate its crystallographic and electronic
    properties, and observe that it has symmetry-protected Dirac
    points at the Fermi level. Its many-body physics is excitingly
    rich, with possible charge, magnetic and superconducting
    instabilities. Through a combination of various many-body methods
    we study possible symmetry-lowering phase transitions such as
    Mott-Hubbard, charge or magnetic ordering, and unconventional
    superconductivity, which in this compound assumes an $f$-wave
    symmetry.}

\end{abstract}

\maketitle








\textbf{Introduction. }It is well known theoretically that graphene is
not the only system where crystallography, combined with a particular
one-electron Hamiltonian, creates symmetry-protected Dirac points at a
certain filling. As such, the hope persists to combine features of
Dirac fermions and strong correlations in an experimentally accessible
condensed matter system. For instance, the single-orbital kagome
tight-binding model is known to feature, besides a flat band that has
been subject of considerable interest, symmetry-protected Dirac points
at a filling of $n=4/3$ electrons per site (or, equivalently, $1/3,$
for reversed sign of the hopping). However, so far this fact has been
considered more a numerical curiosity of a simplified model than an
accessible feature in real materials, and has received little
attention. Currently, the predominantly studied kagome material has
been herbertsmithite~\cite{Shores2005,Lee2008,Mendels2010},
ZnCu$_{3}$(OH)$_{6}$Cl$_{2}$, presumed to host a spin liquid with
fractionalized spin excitations, consistent with recent neutron
scattering measurements~\cite{herbert-nature}. It might be the first
manifestation of a truly two-dimensional resonating valence bond (RVB)
spin liquid anticipated in theory 27 years
ago~\cite{ANDERSON06031987,PhysRevB.35.8865,PhysRevB.44.2664,klein1991resonating},
which is supported by large-scale numerical
investigations~\cite{Yan03062011,PhysRevLett.109.067201,PhysRevB.89.020408}.

\begin{figure}[tbh]
\includegraphics[width=0.5\textwidth]{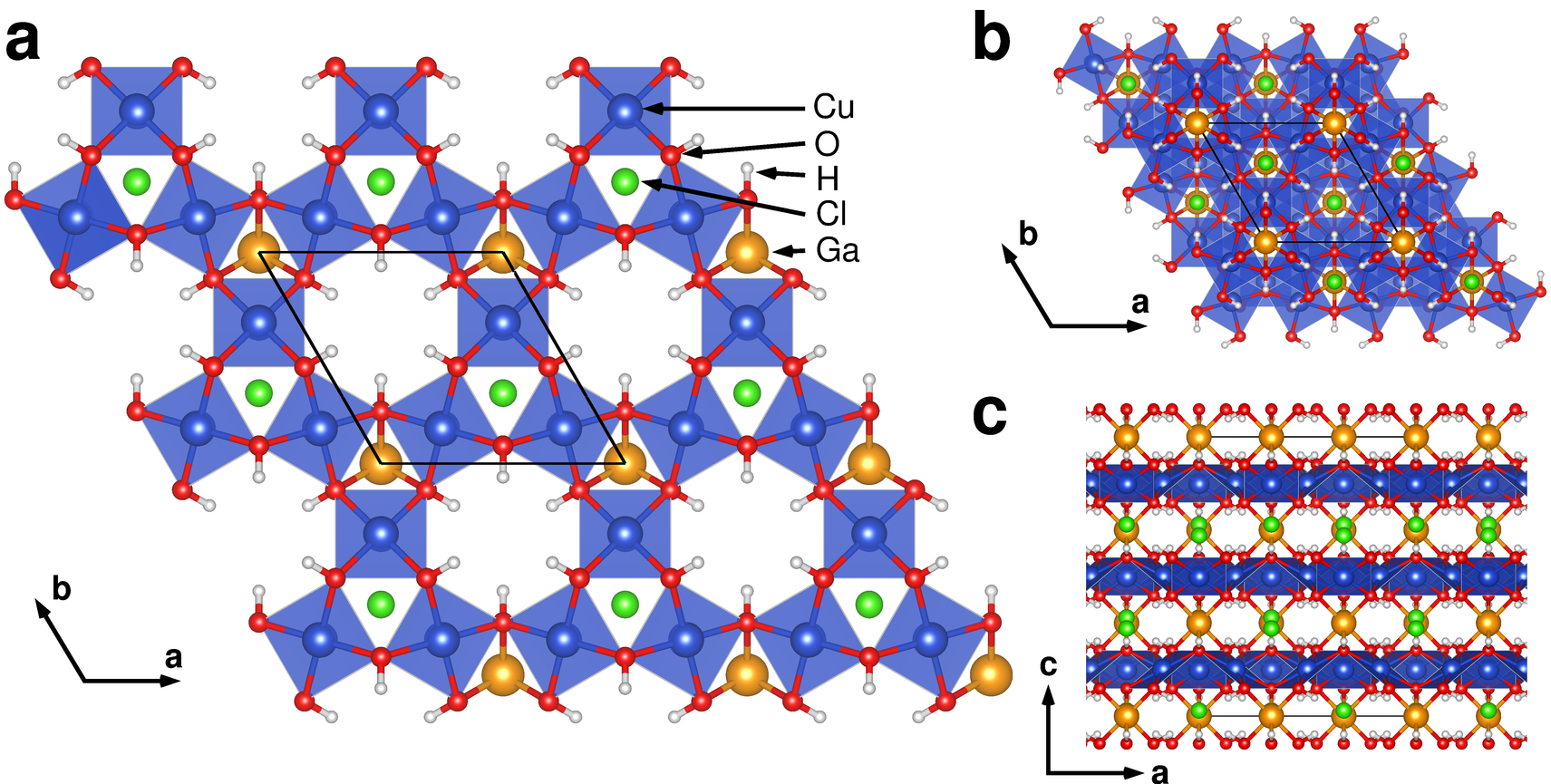}
\caption{\sffamily Structure of predicted herbertsmithite modification
  {\GHS}. \textbf{a} View of an individual kagome
  plane. \textbf{b} Full view along $c$ axis where three shifted
  kagome layers are stacked.  \textbf{c} Side view. }
\label{fig:structure}
\end{figure}

Herbertsmithite~\cite{Jeschke2013}  is the starting point of our proposal to unify strong
correlations, metallicity and Dirac fermions. As opposed to the flat
band which only exists in the nearest neighbour single-orbital tight
binding model, the Dirac points in the kagome compounds are
symmetry-protected and survive in any electronic structure model that
respects the hexagonal symmetry. By replacing Zn$^{2+}$ by Ga$^{3+}$
in herbertsmithite, {\GHS} (Fig.~\ref{fig:structure}), the Dirac
points are placed at the Fermi level. These states are formed by the
strongly correlated ($U/t\sim50$) Cu$-d_{x^{2}-y^{2}}$ orbitals, whose
nearest neighbour hopping Hamiltonian on the kagome lattice appears to
be equivalent to that of $s$-orbitals. As we will further elaborate on
below, we find a strong, compared to graphene, screening, hinting at
rather local interactions, while the nature of magnetic interactions
is typical for the kagome lattice, suggesting no long range
antiferromagnetism and a flat spin fluctuation profile.

Zn and Ga are neighbours in the periodic table and their radii are
similar.  Even if they do not form a continuous solid solution, our
calculations suggest that the fully substituted compound, \GHS, should
be dynamically stable, and it should be possible to dope it with holes
by partially replacing Ga with Zn. If the solid solution
Zn$_{1-x}$Ga$_x$Cu$_{3}$(OH)$_{6}$Cl$_{2}$ were to exist, it would
span the whole range from a fully frustrated RVB spin liquid ($x=0$)
at half filling $n=1$ to a strongly correlated Dirac metal ($x=1$) at
$n=4/3$.  The ground state in the former limit is known to be a
uniform Mott insulator.  Upon doping, a number of instabilities can
manifest themselves, and in fact one can anticipate a complex and
highly interesting phase diagram.  Specifically, the DFT
noninteracting one-electron picture may be prone to (1) Mott-Hubbard
metal-insulator transition, (2) charge ordering, (3) ferromagnetism or
(4) a superconducting instability. Below we will first present our DFT
calculations, and then discuss these possible instabilities by means
of various complementary many-body model approaches.

\textbf{First principles calculations.} Herbertsmithite is part of the
Zn-paratacamite family of compounds~\cite{Mendels2010} that shows
great flexibility in its composition both naturally (\textit{e.g.,}
Ni-herbertsmithite~\cite{Clissold2007}) and synthetically
(\textit{e.g.} Mg-herbertsmithite~\cite{Colman2011}).  While Ga more
commonly assumes a fourfold coordination, it also occurs in the
sixfold one~\cite{Donnay_1971}. In the framework of DFT we fully
relaxed the herbertsmithite structure after replacing Zn by Ga (see
the Supplementary Information).

\begin{figure*}[tbh]
\centering
\includegraphics[width=0.8\textwidth]{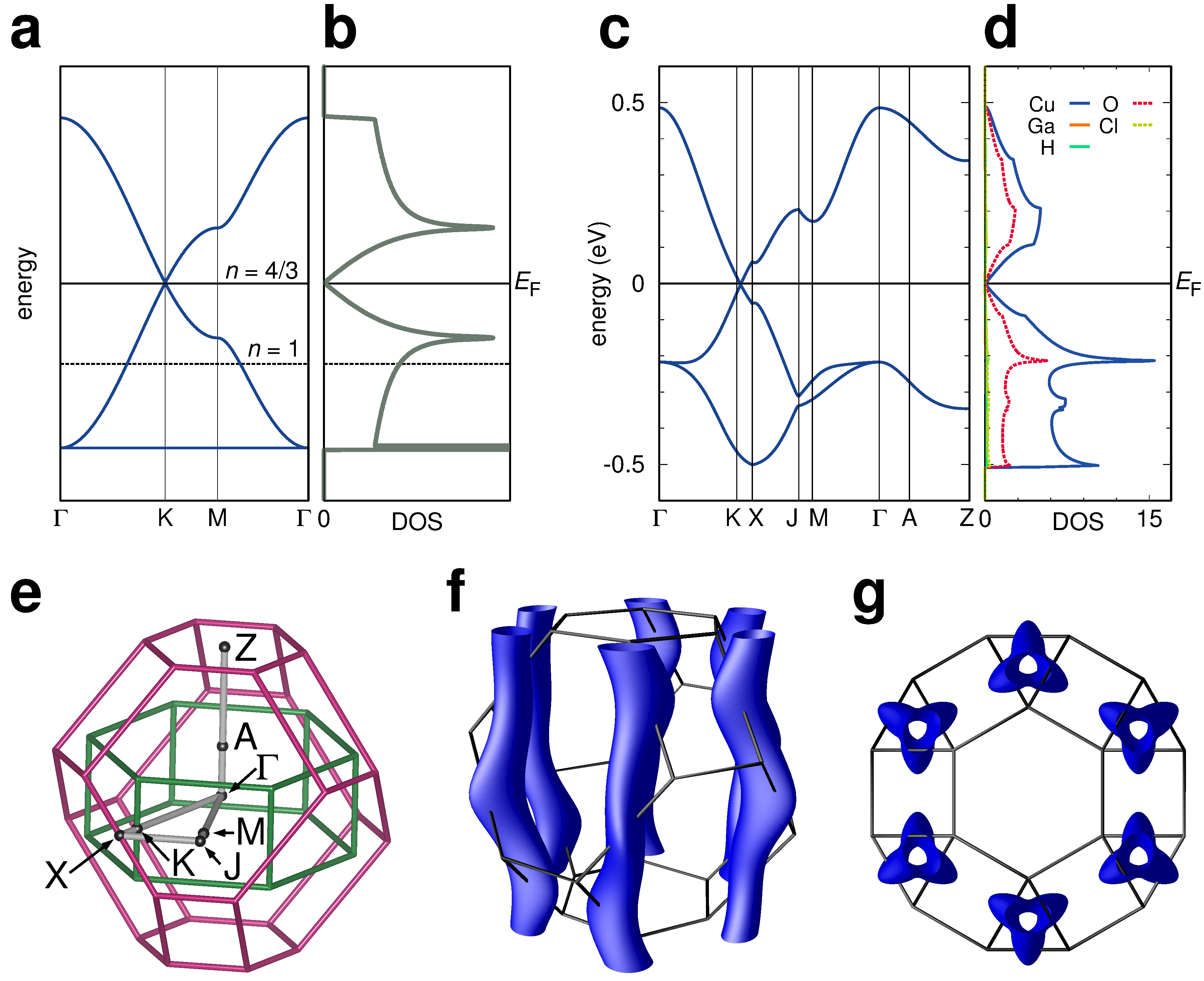}
\caption{\sffamily Electronic structure of {\GHS}.
{\bf a} Band structure and \textbf{b} density of states of
the nearest-neighbor tight-binding model for the kagome lattice.
\textbf{c}-\textbf{g} Electronic structure of Ga-substituted herbertsmithite,
\GHS. \textbf{c}
Band structure along high symmetry points of the hexagonal setting of the
$R\bar{3}m$ space group. \textbf{d} Density of states showing
\GHS to be a zero gap semiconductor. \textbf{e} Brillouin zones
of the $R\bar{3}m$ space group in rhombohedral (purple) and hexagonal setting
(green), with the path chosen in \textbf{c} (see also a view from
the top in Fig.~S1, Supplementary Information). \textbf{f} and \textbf{g} Fermi
surfaces at an energy $E=-60$~meV. The Fermi surface plots
 demonstrate that the material is two-dimensional to a good
approximation. }
\label{fig:electronicstructure}
\end{figure*}

\begin{figure}[tbh]
\includegraphics[width=0.48\textwidth]{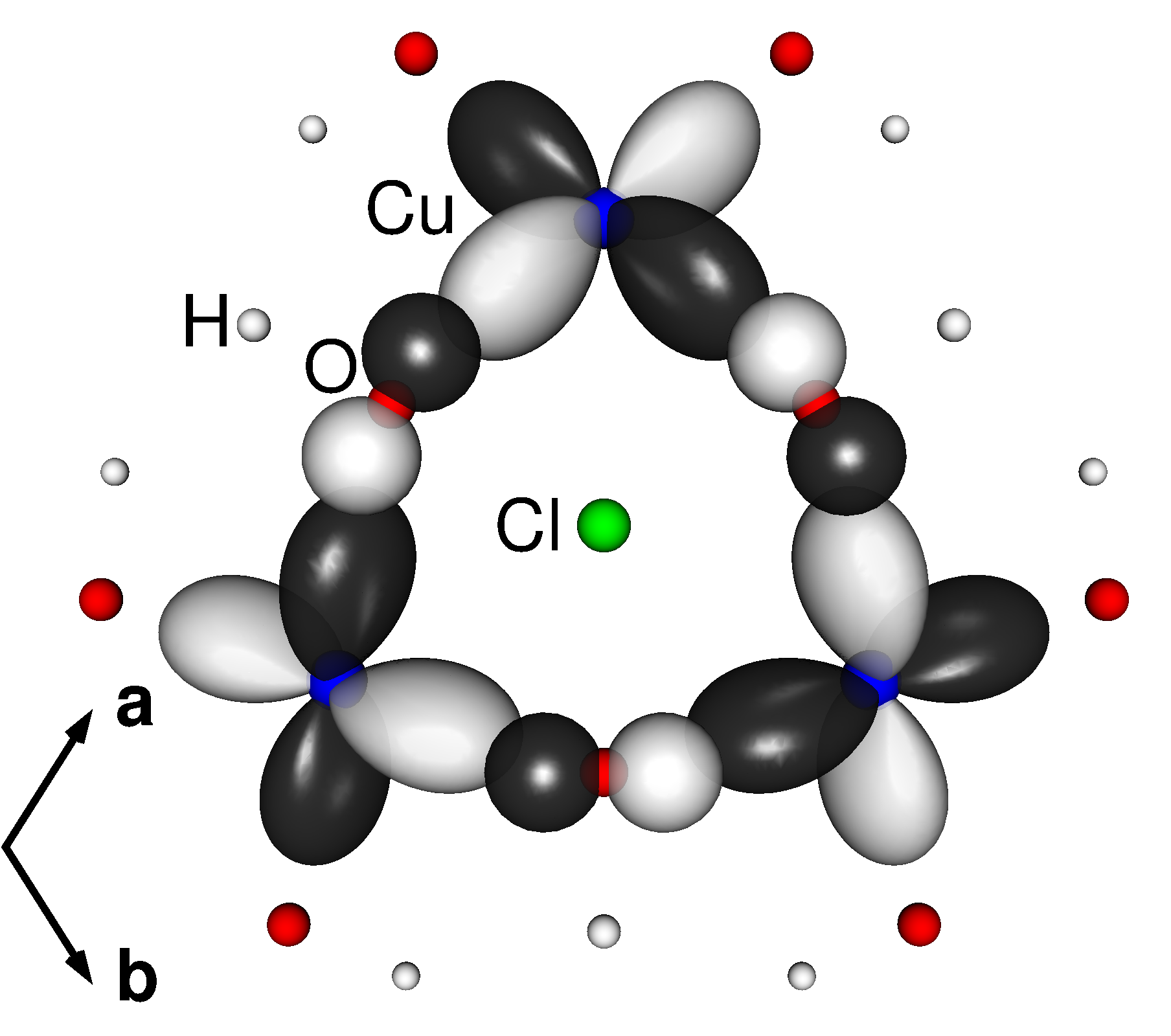}
\caption{\sffamily Relevant tight-binding orbitals
for {\GHS}. Overlap between Cu $3d_{x^{2}-y^{2}}$ and O
$p$ orbitals in a Cu triangle. $+$ and $-$ sign of the wave function is coded
with black and white, respectively.}
\label{fig:orbitals}
\end{figure}

The resulting electronic structure (Fig.~\ref{fig:electronicstructure}
b) shows low-energy bands qualitatively similar to a one-orbital model
on the kagome lattice (Fig.~\ref{fig:electronicstructure} a) and has
Dirac points exactly at the Fermi level. In order to understand this
fact, we first observe that Cu in this structure sits inside an oxygen
square, so that the $d_{x^{2}-y^{2}}$ orbitals, with strong $pd\sigma$
hopping to the oxygens, form a high-lying antibonding band
(Fig.~\ref{fig:orbitals}). If one neglects the other $d$ orbitals, and
integrates out oxygens, one gets a tight-binding model, which is
mathematically equivalent to a single $s$-orbital model with the
nearest neighbour hopping $t=t_{pd\sigma}^{2}/(E_{F}-E_{p})>0.$ The
solution of this model is well known
(Fig.~\ref{fig:electronicstructure}~{\bf a}).  The bands crossing along the
$\Gamma-K$ line have different parity with respect to the $y/-y$
transformation ($y$ $\perp\Gamma K$) , and therefore the Dirac points
at K are symmetry-protected for an arbitrary hopping range (apart from
the very small spin-orbit coupling).

The kagome tight-binding model has Dirac points at $K$; however, this
is not the case for a finite $k_{z}$ dispersion and rhombohedral
symmetry (Fig.~\ref{fig:electronicstructure}~{\bf c}). In a planar system with hexagonal symmetry
one can travel from $\Gamma$ to $M$ and then continue to another $\Gamma$. In
a rhombohedral structure, if one travels along the same direction one actually
ends up at Z, and the $k_z=0$ plane alone has lower symmetry than if averaged 
over all $k_z$.
The Dirac points remain nevertheless protected, but form
a 3D line meandering about the vertical line passing through the 
hexagonal K point (see
Fig.~S1 in the Supplementary Information), which, if averaged over
$k_{z}$, projects onto this point. When doped, the Fermi surface
consists of six tubes that slightly twist as we move along $k_{z}$
(Fig.~\ref{fig:electronicstructure} {\bf f}, {\bf g}).

\textbf{Normal state instabilities.}\textit{-} Next we want to check the system
against normal state electronic instabilities. In a strongly
correlated system this cannot be adequately addressed within the DFT.
Therefore, we have constructed a many-body model Hamiltonian that
captures the most important interactions:
\begin{equation}
\begin{split}
\mathcal{H} &  =\mathcal{H}_{0}+\mathcal{H}_{\text{int}}\\
\mathcal{H}_{0} &  =\sum_{i,j}\sum_{\sigma}t_{ij}c_{i,\sigma}^{\dagger
}c_{j,\sigma}+\mu\sum_{i,\sigma}c_{i,\sigma}^{\dagger}c_{i,\sigma}\\
\mathcal{H}_{\text{int}} &  =U\sum_{i}n_{i,\uparrow}n_{i,\downarrow}+\frac
{V}{2}\sum_{<i,j>}\sum_{\sigma,\sigma^{\prime}}n_{i,\sigma}n_{j,\sigma
^{\prime}}.
\end{split}
\label{hamil}
\end{equation}
where the first term is the effective Cu $d_{x^{2}-y^{2}}$
tight-binding Hamiltonian derived from the DFT band structure (see the
Supplementary Information) and the onsite $U$ and nearest-neighbor
Coulomb repulsion $V$ between Cu $d_{x^{2}-y^{2}}$ atoms we estimated
(see the Supplementary Information) to be $U=5-7$ eV and
$V=0.11$eV. It is worth noting that the latter number is unexpectedly
small. The reason for that is twofold: first, even at the exact
$n=4/3$ filling, where the metallic density of states (DOS) is zero,
the behaviour of the dielectric constant is nearly metallic, and the
Coulomb interaction is well screened; second, even though the
electronic structure is quasi-2D, it is a far cry from an isolated
plane as in graphene --- each plane participates in screening
interactions not only within the plane, but in all other planes.

\textit{Mott-Hubbard instability.-} ZnCu$_{3}$(OH)$_{6}$Cl$_{2}$ is a
Mott insulator since the Cu $d_{x^{2}-y^{2}}$ state is half-filled and
electron hopping incurs an energy cost of $U.$ This is not the case in
{\GHS}. Its filling of $n=4/3$ corresponds to one electron per site
plus one extra electron per three sites. Suppose $U=\infty.$ Then
exactly 1/3 of all sites are double, and the rest are single
occupied. Thus, one electron per site will be localised and 1/3 of an
electron mobile, and this mobility will not be impeded at all. This
simplistic treatment suggests that the system will be metallic.

\begin{figure}[tbh]
\includegraphics[width=0.48\textwidth]{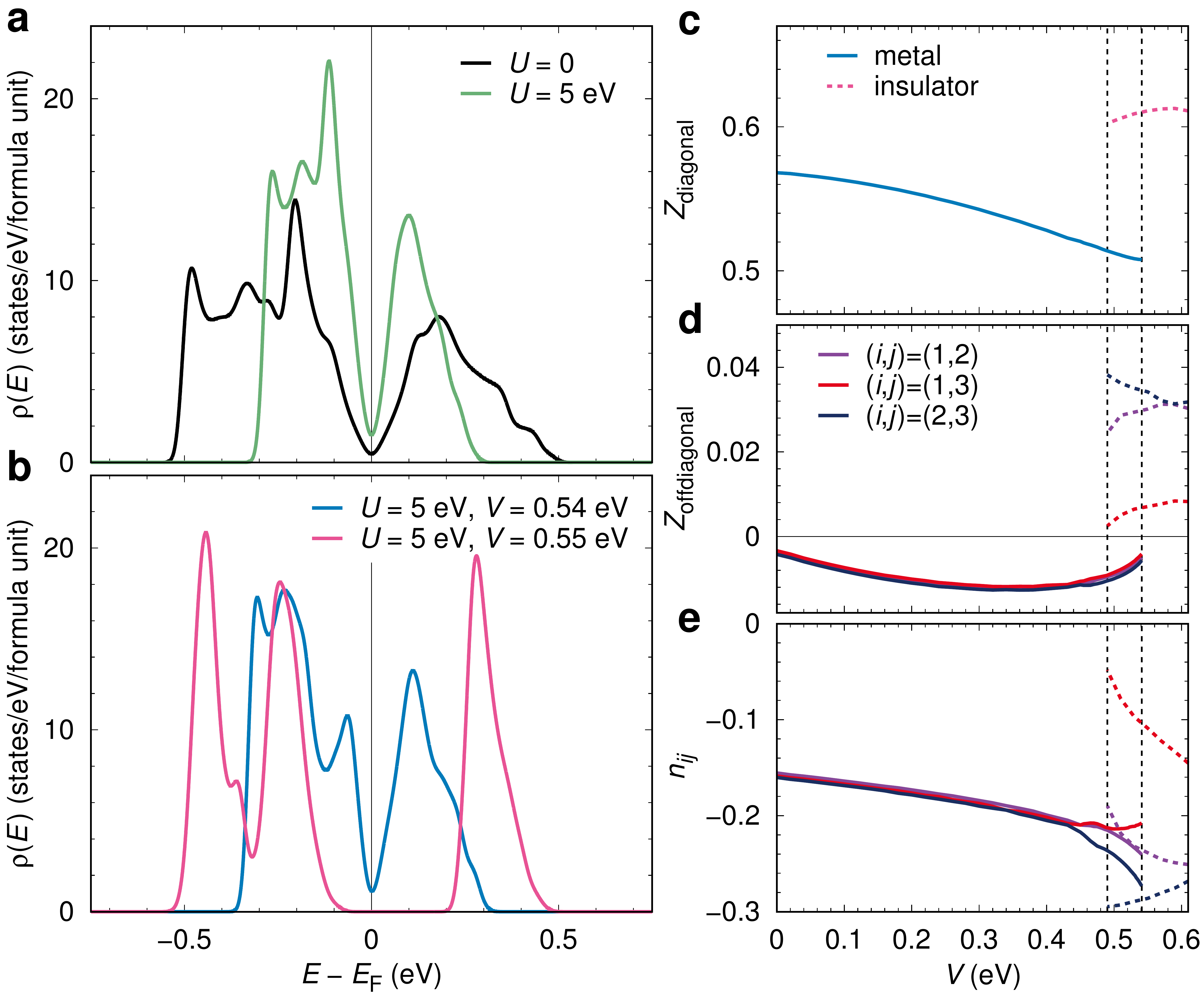}
\caption{\sffamily DCA-like rotationally invariant slave boson results
  for {\GHS}. \textbf{a}-\textbf{b} Quasi-particle (QP) spectral
  function for different interactions on the three-site cluster. The
  curves are scaled with the respective onsite QP
  weight$Z$. \textbf{c}-\textbf{e} cluster quantities with intersite
  $V$: onsite QP weight \textbf{c}, intersite QP weight \textbf{d},
  and intersite terms of the cluster orbital density matrix
  \textbf{e}.}
\label{fig:sb-result}
\end{figure}

To verify that, we applied to the Hamiltonian Eq.~\ref{hamil} the
dynamical cluster approximation (DCA) within the rotationally
invariant slave-boson, DCA(RISB), formalism~\cite{li89,lec07} in the
saddle-point approximation (see the Supplementary Information). As a
minimal cluster we considered three sites.  Figure~\ref{fig:sb-result}
shows our DCA(RISB) results in the paramagnetic regime. For $U=5$~eV, we observe a
band-narrowing but the essential low energy features, especially the
pseudogap reflecting the Dirac points, remain. Thus, the onsite
repulsion alone is insufficient to drive the system insulating at
$n=4/3$. This is a key result, and we have confirmed it by using the
more accurate continuous--time quantum Monte Carlo approach (see the
Supplementary Information).

\textit{Charge ordering.-} Having verified the absence of a
Mott-Hubbard transition, we turn to possible charge ordering, which in
the limit $U\gg t$ is controlled by the intersite Coulomb repulsion
$V$. At the mean-field level, the competition is between the energy
gain of charge ordering, and the loss of kinetic energy. The former
can be estimated from the fact that in the charge ordered state no
neighbouring sites are doubly occupied, thus, the nearest neighbour
repulsion is completely avoided. On the other hand, if all electrons
are randomly distributed over all sites, each nearest neighbour bond
has (1/3)$^{2}$=1/9 probability to have double occupancy on both
sides, losing $V/9$ per bond, or $2V/9$ per site. The kinetic energy
loss can be estimated as ($2/3-1/6)=1/2$ of the energy of
non-interacting electrons on the kagome lattice ($E_{0}\approx-0.8t$
per site for $n=4/3$),~\cite{comment_V}.  Comparing $2V/9$ and $0.4t$
we conclude that the system may develop charge ordering at
$V\gtrsim2t.$ Our DFT calculations yield $t\sim0.3$ eV, however, one
may expect a renormalisation due to correlation effects; a factor of
two is reasonable for Cu $d$ electrons and we get $V_{c}\sim t\sim0.3$
eV. This value is much larger than $V$ estimated for {\GHS}, and,
therefore, it should remain a uniform metal.

Again, we subjected these qualitative arguments to a numerical test
solving Eq.~\ref{hamil} within the DCA(RISB) for $U=5$ eV and
different $V$ values. At a critical $V_{c}=0.54$~eV,
(Fig.~\ref{fig:sb-result}~{\bf b},{\bf c}) well above the estimated
value of $V\approx0.11$ eV, the system charge orders and loses its
metallicity, consistent with the qualitative arguments
above. Interestingly, close to and inside the insulating phase the
system shows substantially increased values of the intersite terms in
the quasi-particle weight (Fig.~\ref{fig:sb-result}~{\bf d}) as well as in
the cluster orbital density matrix (Fig.~\ref{fig:sb-result}~{\bf e}).
Inspection of the slave-boson multiplet amplitudes reveal an insulator
composed of dominant coherent superpositions of the four-particle
triangular-cluster states with one double-occupied site and two
single-occupied sites with different spin orientation, a reflection of
proximity to charge ordering at $V_{c}=0.54$~eV.

Note that in our functional renormalisation group (fRG) calculations,
described below, the problem of the phase instabilities is approached
from the opposite limit, namely the itinerant/weak coupling limit,
valid at $U\lesssim t$, appropriate for addressing the
superconductivity. It is worth mentioning though, that in fRG the
$V_{c}$ needed to stabilise the leading charge ordering instability
was also far greater than the realistic range of $V$.

\textit{Ferromagnetism.-} We consider next the possibility of a
ferromagnetic instability. It is well known that Hubbard models with
very large $U/t$ and away from half-filling show propensity to
so-called Nagaoka ferromagnetism~\cite{Nagaoka}.  Existing
calculations\cite{Hanisch97} suggest that {\GHS} is far from the
Nagaoka regime. However, these works did not consider states with
partial spin-polarization, which are natural in a metallic system.

Let us first estimate the effective magnetic interactions at the
mean-field level by comparing the superexchange
$J_{ex}\sim4t^{2}/U\sim70$ meV and the ferromagnetic kinetic energy
gain per bond, times the number of mobile electrons per site.
$E_{KE}=E_{0}/3\times1/3\sim30$ meV.  $E_{KE}<J_{ex}$, suggesting a
spin-liquid without a long range order. However, this estimate also
does not account for partial polarization, and in fact shows that the
two competing interactions are of the same order of
magnitude. Therefore we looked for ferromagnetism with the DCA(RISB),
and found no ferromagnetism at $n=4/3$ (Dirac metal) and
$n\lesssim4/3$ (hole-doping) for any reasonable parameters.  In
contrast, with electron doping ($n>4/3$), as the superexchange is
gradually suppressed, ferromagnetism becomes increasingly more
favourable.

The importance of electron itineracy is particularly clear from our
weak coupling fRG calculations (see the Supplementary Information). By
construction, fRG shows no instabilities for zero DOS, but as we dope
away from the Dirac point (either holes or electrons), we find, at
sufficiently large $U$, a ferromagnetic instability. It should be
noted that fRG depends on resummation of all parquet diagrams, an
approximation not rigorously justified when $U$ exceeds the band
width. As we do not see clean divergent RG flow for the ferromagnetic
channel, we cannot distinguish ferromagnetic fluctuations from a
partially-polarised ferromagnet. From the combined view of different
methodolgies, however, we can ascertain that the system is very close
to a ferromagnetic instability; whether it is realized in {\GHS}, we
cannot be sure.  But, importantly, strong ferromagnetic fluctuations
are essential for unconventional superconductivity, discussed in the
next section.

\textbf{Superconductivity.}\textit{-} Finally, we investigate the
possibility of superconductivity in {\GHS}. Obviously, when the Fermi
level is exactly at the Dirac point the effective DOS (averaged over
the cutoff energy) is small, of the order of $J_{ex}v_{F},$ where the
cutoff is set at the exchange energy; $v_{F}$ is the Dirac
velocity~\cite{lambda}.  However if the Ga substitution is incomplete
and the Fermi surfaces form narrow cylinders
(Fig.~\ref{fig:electronicstructure}~{\bf f},{\bf g}) one expects sizeable coupling
with spin excitations. This can be qualitatively estimated in the
simple model of Fig.~\ref{fig:electronicstructure}~{\bf a} as follows: The
pairing interaction can be written, roughly, as $t^{2}S(q,\omega),$
where $S(q,\omega)$ is the Fourier-transform of the spin-spin
correlator $\left\langle S_{i}S_{j}\right\rangle .$ Assuming that only
the nearest spins are correlated (antiferromagnetically), we get three
branches, one corresponding to an uncorrelated spin liquid,
$S(q,\omega)=0,$ and two dispersive ones. As discussed in
Ref.~\onlinecite{PhysRevB.45.2899}, this corresponds to three magnon
modes, one dispersionless and two acoustic; the corresponding pairing
interaction is strong at $q\sim0$ and weakens for larger $q$ values, and is
repulsive for singlet and attractive for triplet pairing. Note that
possible ferromagnetic fluctuations, not included in this simplistic
picture, are $also$ peaked at $q\sim0$ and thus the arguments above
apply to them, too.

Now we observe that in a hexagonal Brillouin zone the 6 K-points
related by the rotational symmetry (Supplementary Information,
Fig.~S1) can be separated into two subsets, $K$ and $K^{\prime},$ such
that all points of a given subset are also connected by translational
symmetry.  Given that the superconducting order parameter has to honor
translational symmetry (but not necessarily rotational one), we see
that as we go around the Brillouin zone the phase factors between the
consecutive pockets (neglecting intrapocket variations) vary as
$\phi,$ $\phi^{\prime},$ $\phi,$ $\phi^{\prime},$ $\phi,$
$\phi^{\prime },$ corresponding to an $L=3$ ($f$-wave)
harmonic. Indeed there exists a state~\cite{RevModPhys.63.239}, often
labelled $B_{1u}$, with the order parameter transforming as
$\mathbf{\hat{z}}x(x^{2}-3y^{2}),$ so that $\phi=-\phi^{\prime}.$ In
this case the intrapocket ($e.g.$, small $q$) interaction is
attractive and pairing, while the interpocket one is repulsive and
pairbreaking. However, the interpocket interaction is suppressed
(assuming an acoustic spectrum) as ($k_{F}/G)^{2}$, with the distance
$G=K-K^{\prime}$. One can estimate the BCS $T_{c}$=$\omega
\exp(-1/\lambda),$ where $\omega\sim Jk_{F}a$
(Ref.~\onlinecite{PhysRevB.45.2899}, Eq. 2.23), and $\lambda\sim
t^{2}%
N(E_{F})/\omega\sim t/J.$ We see that (i) the coupling constant is
roughly independent of doping, (ii) the prefactor changes
approximately as a square root of doping and (iii) one can expect a
sizeable $T_{c}$ for $k_{F}\lesssim G/2$ (the maximum possible value
before changing the topology). For $\lambda\sim1,$ which seems to be a
conservative estimate, and $k_{F}\sim G/4,$ we get
$T_{c}\sim0.07J\sim30$ K for $J\sim40$ meV. Note that in this regime
dependence on $\lambda$ is close to linear. Taking $\lambda\sim2$
raises the estimate for $T_{c}$ to 60 K. Of course, all these
estimates are order of magnitude at best, therefore we turn again to a
quantitative treatment in terms of multi-sublattice
fRG~\cite{pht}. Note that superconductivity is an itinerant effect,
and the inherent fRG assumption of weak coupling should be
acceptable.

Around $n=4/3$ and constraining ourselves to the superconducting
channel, which, from a Kohn-Luttinger perspective, would become the
leading channel for sufficiently weak interactions, we find a clear
preference for nodeless $f$-wave superconductivity, as anticipated
from the qualitative argument above. The calculated gaps, while
changing sign between the pockets, remain uniform inside each pocket
(see Supplementary Information, Fig.~S3). It is worth noting that the
disorder introduced by Zn doping, while weak, might even,
counterintuitively, support superconductivity in this
limit\cite{PhysRevB.87.174511}. Finally, a subtlety could arise for
significant $V$: In order to avoid onsite \textit{and} nearest
neighbour repulsion, this could drive a transition from a nodeless
$B_{1u}$ $f$-wave to nodal $B_{2u}$ $f$-wave~\cite{Ronny}, but for
realistic values of $V$ this is highly unlikely.

{\bf Conclusions.}\textit{-} This work opens many interesting avenues
toward novel physics in electron-doped herbertsmithite
compounds. Bridging between an RVB scenario of
frustrated magnetism for $x=0$ and a correlated Dirac metal at $x=1$,
$\text{Zn}_{1-x}\text{Ga}_{x}$Cu$_{3}$(OH)$_{6}$Cl$_{2}$ promises to
exhibit a plethora of unconventional electronic phases as a function
of doping, temperature, and disorder. Even if a solid solution $0<x<1$
(which would be an ultimate bonanza of novel physics) would not be
realized experimentally, pure {\GHS} ($x=1)$ and its slightly hole
doped version ($x\lesssim1)$ should be highly interesting systems, and
can still be characterized as doped RVB. According to our analysis, in
this regime (close to the Dirac points) the two leading candidates for
the low-temperature ground state are weak ferromagnetism and $f$-wave
superconductivity. Further experimental and theory work shall decide
who is the winner in their competition.

{\bf Acknowledgements}

We acknowledge useful discussions with C. Krellner, C. Platt, G. Khalliulin,
A. Chubukov and C. Piefke. I.I.M. is supported by ONR through the NRL
basic research program.  H.O.J., F.L. and R.V. are supported by 
DFG-FOR1346 and DFG-SFB/TR49 (H.O.J., R.V.). R.T. is supported by the
European Research Council through the grant TOPOLECTRICS,
ERC-StG-336012.

{\bf Author contributions}
 Ab initio density functional 
 calculations, conceptual development and analysis: I.I.M., H.O.J., R.V.
DCA(RISB) calculations: F.L. 
DCA(CT-QMC) calculations H.L.
fRG calculations: M.F, R.T.
All authors participated in writing the manuscript.


\begin{thebibliography}{10}
\expandafter\ifx\csname url\endcsname\relax
  \def\url#1{\texttt{#1}}\fi
\expandafter\ifx\csname urlprefix\endcsname\relax\def\urlprefix{URL }\fi
\providecommand{\bibinfo}[2]{#2}
\providecommand{\eprint}[2][]{\url{#2}}

\bibitem{Shores2005}
\bibinfo{author}{Shores, M.~P.}, \bibinfo{author}{Nytko, E.~A.},
  \bibinfo{author}{Bartlett, B.~M.} \& \bibinfo{author}{Nocera, D.}
\newblock \bibinfo{title}{A structurally perfect \mbox{$S=1/2$} kagome
  antiferromagnet}.
\newblock \emph{\bibinfo{journal}{J. Am. Chem. Soc.}}
  \textbf{\bibinfo{volume}{127}}, \bibinfo{pages}{13462}
  (\bibinfo{year}{2005}).

\bibitem{Lee2008}
\bibinfo{author}{Lee, P.~A.}
\newblock \bibinfo{title}{An end to the drought of quantum spin liquids}.
\newblock \emph{\bibinfo{journal}{Science}} \textbf{\bibinfo{volume}{321}},
  \bibinfo{pages}{1306--1307} (\bibinfo{year}{2008}).
\newblock
\newblock 

\bibitem{Mendels2010}
\bibinfo{author}{Mendels, P.} \& \bibinfo{author}{Bert, F.}
\newblock \bibinfo{title}{Quantum kagome antiferromagnet
  \mbox{ZnCu$_3$(OH)$_6$Cl$_2$}}.
\newblock \emph{\bibinfo{journal}{J. Phys. Soc. Jpn.}}
  \textbf{\bibinfo{volume}{79}}, \bibinfo{pages}{011001}
  (\bibinfo{year}{2010}).

\bibitem{herbert-nature}
\bibinfo{author}{Han, T.-H.} \emph{et~al.}
\newblock \bibinfo{title}{Fractionalized excitations in the spin-liquid state
  of a kagome-lattice antiferromagnet}.
\newblock \emph{\bibinfo{journal}{Nature}} \textbf{\bibinfo{volume}{492}},
  \bibinfo{pages}{406} (\bibinfo{year}{2012}).

\bibitem{ANDERSON06031987}
\bibinfo{author}{Anderson, P.~W.}
\newblock \bibinfo{title}{The resonating valence bond state in
  \mbox{La$_2$CuO$_4$} and superconductivity}.
\newblock \emph{\bibinfo{journal}{Science}} \textbf{\bibinfo{volume}{235}},
  \bibinfo{pages}{1196--1198} (\bibinfo{year}{1987}).
\newblock
\newblock 

\bibitem{PhysRevB.35.8865}
\bibinfo{author}{Kivelson, S.~A.}, \bibinfo{author}{Rokhsar, D.~S.} \&
  \bibinfo{author}{Sethna, J.~P.}
\newblock \bibinfo{title}{Topology of the resonating valence-bond state:
  Solitons and high \mbox{$T_c$} superconductivity}.
\newblock \emph{\bibinfo{journal}{Phys. Rev. B}} \textbf{\bibinfo{volume}{35}},
  \bibinfo{pages}{8865--8868} (\bibinfo{year}{1987}).
\newblock 

\bibitem{PhysRevB.44.2664}
\bibinfo{author}{Wen, X.~G.}
\newblock \bibinfo{title}{Mean-field theory of spin-liquid states with finite
  energy gap and topological orders}.
\newblock \emph{\bibinfo{journal}{Phys. Rev. B}} \textbf{\bibinfo{volume}{44}},
  \bibinfo{pages}{2664--2672} (\bibinfo{year}{1991}).
\newblock 

\bibitem{klein1991resonating}
\bibinfo{author}{Klein, D.}, \bibinfo{author}{Schmalz, T.~G.},
  \bibinfo{author}{Garc{\'\i}a~Bach, M.~A.}, \bibinfo{author}{Valent{\'\i}, R.}
  \& \bibinfo{author}{{\v{Z}}ivkovic, T.~P.}
\newblock \bibinfo{title}{Resonating-valence-bond theory for the square-planar
  lattice}.
\newblock \emph{\bibinfo{journal}{Phys. Rev. B}} \textbf{\bibinfo{volume}{43}}.

\bibitem{Yan03062011}
\bibinfo{author}{Yan, S.}, \bibinfo{author}{Huse, D.~A.} \&
  \bibinfo{author}{White, S.~R.}
\newblock \bibinfo{title}{Spin-liquid ground state of the \mbox{$S = 1/2$}
  kagome Heisenberg antiferromagnet}.
\newblock \emph{\bibinfo{journal}{Science}} \textbf{\bibinfo{volume}{332}},
  \bibinfo{pages}{1173--1176} (\bibinfo{year}{2011}).
\newblock
\newblock 

\bibitem{PhysRevLett.109.067201}
\bibinfo{author}{Depenbrock, S.}, \bibinfo{author}{McCulloch, I.~P.} \&
  \bibinfo{author}{Schollw\"ock, U.}
\newblock \bibinfo{title}{Nature of the spin-liquid ground state of the
  \mbox{$S=1/2$} Heisenberg model on the kagome lattice}.
\newblock \emph{\bibinfo{journal}{Phys. Rev. Lett.}}
  \textbf{\bibinfo{volume}{109}}, \bibinfo{pages}{067201}
  (\bibinfo{year}{2012}).
\newblock

\bibitem{PhysRevB.89.020408}
\bibinfo{author}{Suttner, R.}, \bibinfo{author}{Platt, C.},
  \bibinfo{author}{Reuther, J.} \& \bibinfo{author}{Thomale, R.}
\newblock \bibinfo{title}{Renormalization group analysis of competing quantum
  phases in the \mbox{$J_1$-$J_2$} Heisenberg model on the kagome lattice}.
\newblock \emph{\bibinfo{journal}{Phys. Rev. B}} \textbf{\bibinfo{volume}{89}},
  \bibinfo{pages}{020408} (\bibinfo{year}{2014}).
\newblock 

\bibitem{Jeschke2013}
\bibinfo{author}{Jeschke, H. O.}, \bibinfo{author}{Salvat-Pujol, F.},
\& \bibinfo{author}{Valent\'\i, R.}
\newblock \bibinfo{title}{First-principles determination of Heisenberg Hamiltonian parameters for
the spin-1/2 kagome antiferromagnet ZnCu$_3$(OH)$_6$Cl$_2$}.
\newblock \emph{\bibinfo{journal}{Phys. Rev. B}} \textbf{\bibinfo{volume}{88}},
  \bibinfo{pages}{075106} (\bibinfo{year}{2013}). 

\bibitem{Clissold2007}
\bibinfo{author}{Clissold, M.~E.}, \bibinfo{author}{Leverett, P.},
  \bibinfo{author}{Williams, P.~A.}, \bibinfo{author}{Hibbs, D.~E.} \&
  \bibinfo{author}{Nickel, E.~H.}
\newblock \bibinfo{title}{The structure of gillardite, the Ni-analogue of
  herbertsmithite, from Widgiemooltha, Western Australia}.
\newblock \emph{\bibinfo{journal}{Can. Mineralog.}}
  \textbf{\bibinfo{volume}{45}}, \bibinfo{pages}{317} (\bibinfo{year}{2007}).

\bibitem{Colman2011}
\bibinfo{author}{Colman, R.~H.}, \bibinfo{author}{Sinclair, A.} \&
  \bibinfo{author}{Wills, A.~S.}
\newblock \bibinfo{title}{Magnetic and crystallographic studies of
  \mbox{Mg}-herbertsmithite, \mbox{$\gamma$-Cu$_3$Mg(OH)$_6$Cl$_2$} -- a new
  \mbox{$S = 1/2$} kagome magnet and candidate spin liquid}.
\newblock \emph{\bibinfo{journal}{Chem. Mater.}} \textbf{\bibinfo{volume}{23}},
  \bibinfo{pages}{1811} (\bibinfo{year}{2011}).

\bibitem{Donnay_1971}
\bibinfo{author}{Scott, J. D.}
\newblock \bibinfo{title}{Crystal structure of a new mineral, s\"ohngeite}.
\newblock \emph{\bibinfo{journal}{The American Mineralogist}}
  \textbf{\bibinfo{volume}{56}}, \bibinfo{pages}{355} (\bibinfo{year}{1971}).

\bibitem{li89}
\bibinfo{author}{Li, T.}, \bibinfo{author}{W\"olfle, P.} \&
  \bibinfo{author}{Hirschfeld, P.~J.}
\newblock \bibinfo{title}{Spin-rotation-invariant slave-boson approach to the
  hubbard model}.
\newblock \emph{\bibinfo{journal}{Phys. Rev. B}} \textbf{\bibinfo{volume}{40}},
  \bibinfo{pages}{6817--6821} (\bibinfo{year}{1989}).
\newblock 

\bibitem{lec07}
\bibinfo{author}{Lechermann, F.}, \bibinfo{author}{Georges, A.},
  \bibinfo{author}{Kotliar, G.} \& \bibinfo{author}{Parcollet, O.}
\newblock \bibinfo{title}{Rotationally invariant slave-boson formalism and
  momentum dependence of the quasiparticle weight}.
\newblock \emph{\bibinfo{journal}{Phys. Rev. B}} \textbf{\bibinfo{volume}{76}},
  \bibinfo{pages}{155102} (\bibinfo{year}{2007}).
\newblock 

\bibitem{comment_V}
\bibinfo{note}{In the charge-disordered state any electron has a 2/3
  probability that a given neighbouring site is single-occupied and allows
  hopping, while in the ordered state half of the electrons are immobile, and
  the mobile ones have only 1/3 of the nearest neighbours single-occupied.}

\bibitem{Nagaoka}
\bibinfo{author}{Nagaoka, Y.}
\newblock \bibinfo{title}{Ferromagnetism in a narrow, almost half-filled s
  band}.
\newblock \emph{\bibinfo{journal}{Physical Review}}
  \textbf{\bibinfo{volume}{147}}, \bibinfo{pages}{392} (\bibinfo{year}{1966}).

\bibitem{Hanisch97}
\bibinfo{author}{Hanisch, T.}, \bibinfo{author}{Uhrig, G.~S.} \&
  \bibinfo{author}{M\"uller-Hartmann, E.}
\newblock \bibinfo{title}{Lattice dependence of saturated ferromagnetism in the
  hubbard model}.
\newblock \emph{\bibinfo{journal}{Phys. Rev. B}} \textbf{\bibinfo{volume}{56}},
  \bibinfo{pages}{13960--13982} (\bibinfo{year}{1997}).
\newblock 

\bibitem{lambda}
\bibinfo{note}{Not only the coupling constant $\lambda$ is small in this case,
  the critical temperature $T_{c}\propto\exp(-1/\lambda^{2}),$ rather than
  $\exp(-1/\lambda)$}.

\bibitem{PhysRevB.45.2899}
\bibinfo{author}{Harris, A.~B.}, \bibinfo{author}{Kallin, C.} \&
  \bibinfo{author}{Berlinsky, A.~J.}
\newblock \bibinfo{title}{Possible N{\'e}el orderings of the kagome
  antiferromagnet}.
\newblock \emph{\bibinfo{journal}{Phys. Rev. B}} \textbf{\bibinfo{volume}{45}},
  \bibinfo{pages}{2899--2919} (\bibinfo{year}{1992}).
\newblock 

\bibitem{RevModPhys.63.239}
\bibinfo{author}{Sigrist, M.} \& \bibinfo{author}{Ueda, K.}
\newblock \bibinfo{title}{Phenomenological theory of unconventional
  superconductivity}.
\newblock \emph{\bibinfo{journal}{Rev. Mod. Phys.}}
  \textbf{\bibinfo{volume}{63}}, \bibinfo{pages}{239--311}
  (\bibinfo{year}{1991}).
\newblock 

\bibitem{pht}
\bibinfo{author}{Platt, C.}, \bibinfo{author}{Hanke, W.} \&
  \bibinfo{author}{Thomale, R.}
\newblock \bibinfo{title}{Functional renormalization group for multi-orbital
  fermi surface instabilities}.
\newblock \emph{\bibinfo{journal}{Advances in Physics}}
  \textbf{\bibinfo{volume}{62}}, \bibinfo{pages}{453} (\bibinfo{year}{2013}).

\bibitem{PhysRevB.87.174511}
\bibinfo{author}{Nandkishore, R.}, \bibinfo{author}{Maciejko, J.},
  \bibinfo{author}{Huse, D.~A.} \& \bibinfo{author}{Sondhi, S.~L.}
\newblock \bibinfo{title}{Superconductivity of disordered dirac fermions}.
\newblock \emph{\bibinfo{journal}{Phys. Rev. B}} \textbf{\bibinfo{volume}{87}},
  \bibinfo{pages}{174511} (\bibinfo{year}{2013}).
\newblock 

\bibitem{Ronny}
\bibinfo{author}{Kiesel, M.~L.}, \bibinfo{author}{Platt, C.},
  \bibinfo{author}{Hanke, W.}, \bibinfo{author}{Abanin, D.~A.} \&
  \bibinfo{author}{Thomale, R.}
\newblock \bibinfo{title}{Competing many-body instabilities and unconventional
  superconductivity in graphene}.
\newblock \emph{\bibinfo{journal}{Phys. Rev. B}} \textbf{\bibinfo{volume}{86}},
  \bibinfo{pages}{020507} (\bibinfo{year}{2012}).
\newblock 

\end{thebibliography}

\end{document}


\title{Theoretical prediction of a strongly correlated Dirac metal\\-- Supplementary Material --}
\author{I. I. Mazin}
\affiliation{Code 6393, Naval Research Laboratory, Washington, DC 20375, USA}
\author{Harald O. Jeschke}
\affiliation{Institut f\"ur Theoretische Physik, Goethe-Universit\"at Frankfurt,
Max-von-Laue-Strasse 1, 60438 Frankfurt am Main, Germany}
\author{Frank Lechermann}
\affiliation{I. Institut f\"ur Theoretische Physik, Universit\"at Hamburg, D-20355 Hamburg, Germany}
\author{Hunpyo Lee}
\affiliation{Institut f\"ur Theoretische Physik, Goethe-Universit\"at Frankfurt,
Max-von-Laue-Strasse 1, 60438 Frankfurt am Main, Germany}
\author{Mario Fink}
\affiliation{Institut f\"ur Theoretische Physik I, Universit\"at W\"urzburg, am Hubland,
97074 W\"urzburg, Germany}
\author{Ronny Thomale}
\affiliation{Institut f\"ur Theoretische Physik I, Universit\"at W\"urzburg, am Hubland,
97074 W\"urzburg, Germany}
\author{Roser Valent\'\i}
\affiliation{Institut f\"ur Theoretische Physik, Goethe-Universit\"at Frankfurt,
Max-von-Laue-Strasse 1, 60438 Frankfurt am Main, Germany}
\date{\today}

\maketitle

\renewcommand{\thetable}{S\Roman{table}}
\renewcommand{\thefigure}{S\arabic{figure}} \renewcommand{\thesection}{}
\renewcommand{\thesubsection}{S\arabic{subsection}} \setcounter{figure}{0} \setcounter{table}{0}

\subsection{First principles calculations}

For structural optimization and to probe structural stability we used
the projector augmented wave method as implemented in the VASP
package~\cite{Kresse1993,Kresse1996}, with the generalised gradient
approximation (GGA) functional~\cite{GGA}. We show in
Table~\ref{tab:ga} the predicted structure of {\GHS} which can be
compared to the experimental crystal structure of herbertsmithite
(Table~\ref{tab:zn}). Note that the optimized {\GHS} structure is very
close to the herbertsmithite one, indicating that the kagome structure
is as natural for Ga as it is Zn.

We have also verified that this structure is dynamically stable
against $q\rightarrow0$ perturbations by veryfing that the perturbed
structures relax back to the original structure.

For the calculation of the dielectric constant (see below) we used the Linear
Augmented Plane Wave method as implemented in the WIEN2k package~\cite{wien2k}. 

Electronic structure plots of {\GHS}, presented in this work, as well as the
tight binding parameters, were obtained using the full
potential local orbital (FPLO) package~\cite{Koepernik1999} together with the
generalized gradient approximation (GGA)~\cite{GGA} functional. For the
density of states in Fig.~2~\textbf{d} of the main text, we use a very dense
$100\times100\times100$ $k$ mesh. To obtain a tight binding representation of
the Cu $3d_{x^{2}-y^{2}}$ bands near the Fermi level, we project the
(Kohn-Sham) Bloch states onto localized orbitals using projective Wannier
functions within the FPLO basis as described in Ref.~\onlinecite{Eschrig2009}.

\begin{figure}
\includegraphics[width=0.5\textwidth]{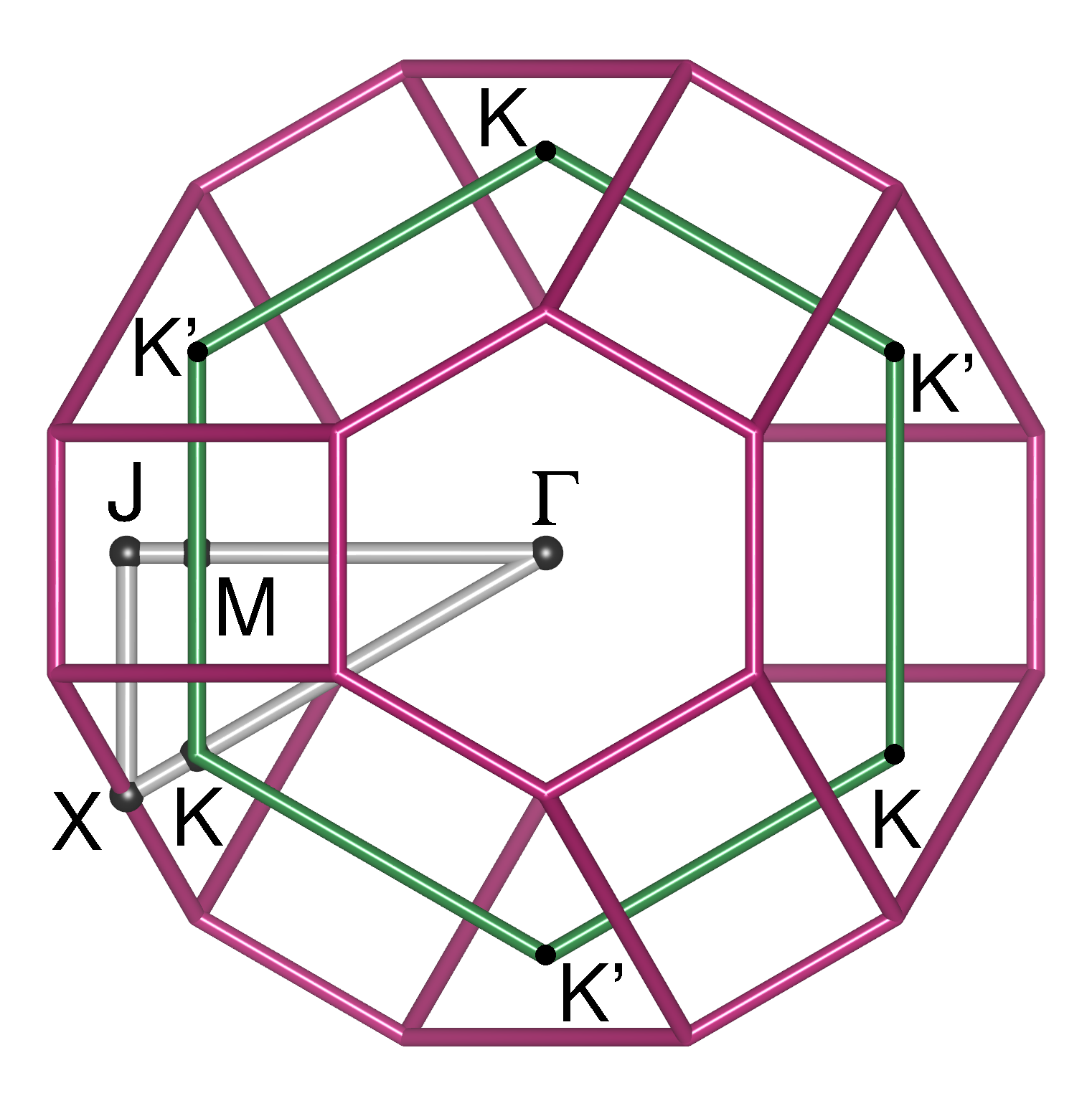}
\caption{\sffamily Brillouin zone of {\GHS} viewed along the $k_z$
  direction. The Brillouin zone of the $R\bar{3}m$ space group is
  shown in rhombohedral (purple) and hexagonal setting (green). The
  path path chosen in Fig. 2 \textbf{c} of the main text is indicated
  in gray.} \label{fig:bz}
\end{figure}

Finally, in Fig.~\ref{fig:bz} we provide an additional view of the Brillouin zone of {\GHS}. 

\begin{table}
\caption{\sffamily Experimental structure of herbertsmithite~\cite{Shores2005}. Space group $R\,\bar{3}m$, $a=6.8342$
~{\AA}, $c=14.0320$~{\AA}, $\alpha=\beta=90^\circ$, $\gamma=120^\circ$. }\label{tab:zn}
\begin{ruledtabular}\sffamily
\begin{tabular}{cD{.}{.}{1.5}D{.}{.}{1.5}D{.}{.}{1.5}}
type& $x$&$y$&$z$\\\hline
Zn&  0 &  0 &  0  \\
Cu&  \nicefrac{1}{3} &  \nicefrac{1}{6} &  \nicefrac{1}{6}  \\
O&   0.1265&   0.2529 &  0.1050  \\
H&   0.192  &  0.384  & 0.084 \\
Cl & 0  & 0 & 0.30521  
\end{tabular}
\end{ruledtabular}
\end{table}

\begin{table}
\caption{\sffamily Predicted structure of {\GHS}. Space group $R\,\bar{3}m$, $a=6.97794$
~{\AA}, $c=13.45890$~{\AA}, $\alpha=\beta=90^\circ$, $\gamma=120^\circ$. }\label{tab:ga}
\begin{ruledtabular}\sffamily
\begin{tabular}{cD{.}{.}{1.5}D{.}{.}{1.5}D{.}{.}{1.5}}
type& $x$&$y$&$z$\\\hline
Ga&  0 &  0 &  0  \\
Cu&  \nicefrac{1}{3} &  \nicefrac{1}{6} &  \nicefrac{1}{6}  \\
O& 0.12441 &0.24882 &0.09852\\
H&0.19846 &0.39692 &0.06555\\
Cl & 0  & 0 & 0.31342  
\end{tabular}
\end{ruledtabular}
\end{table}

\subsection{Determination of U and V for {\GHS}}

The Hubbard repulsion $U$ for copper is large, of the order of 5-7 eV. This
value was obtained by using the method
proposed in Ref.~\onlinecite{PhysRevB.67.153106}. Estimating the intersite
Coulomb repulsion $V$ is more involved. We use here the Thomas-Fermi theory
for semiconductors as formulated by Resta~\cite{Resta} where the screened
potential $V(r)$ of a point charge $e$ is given by:
\begin{equation}
V(r)=\frac{e^{2}}{r}\frac{\sinh\kappa(R-r)}{\sinh\kappa R}+\frac{Ze^{2}%
}{\varepsilon_{0}R}, \label{Resta}%
\end{equation}
$\varepsilon_{0}$ is the static dielectric constant, $\kappa$ is the
Thomas-Fermi vector defined as $\kappa^{2}=4(3\rho/\pi)^{1/3},$ and the
screening parameter $R$ is defined through the relation%
\begin{equation}
\sinh\kappa R/\kappa R=\varepsilon_{0}.
\end{equation}
This relation is valid when $r\lesssim R.$ For the total density of the
valence electrons we can take ($4/3)/\Omega,$ where $\Omega$ is the unit cell
volume, 200 \AA $^{3}$. Note that this way we totally neglect screening by
other electrons but Cu $d_{x^{2}-y^{2}},$ but they probably contribute little.
We calculated $\varepsilon_{0}$  and obtained $\varepsilon_{xx0}=80$
(screening along $z$ is much smaller, about 2.5. For simplicity we will use
simply the in-plane dielectric constant), $r$ here is the Cu-Cu distance, 3.4 \AA .

Solving for $R,$ we verify that $R=11.2 \mathring{A}\gg r.$ Substituting these
values into Eq.~\ref{Resta} we find $V(r)=0.11$ eV~\cite{hadwetaken}. For the
next neighbours, $d_{2}=d_{1}\sqrt{3},$ $V_{2}=0.03\ll V$.

\subsection{DCA(CT-QMC) details}

\begin{figure}
\includegraphics[width=0.5\textwidth]{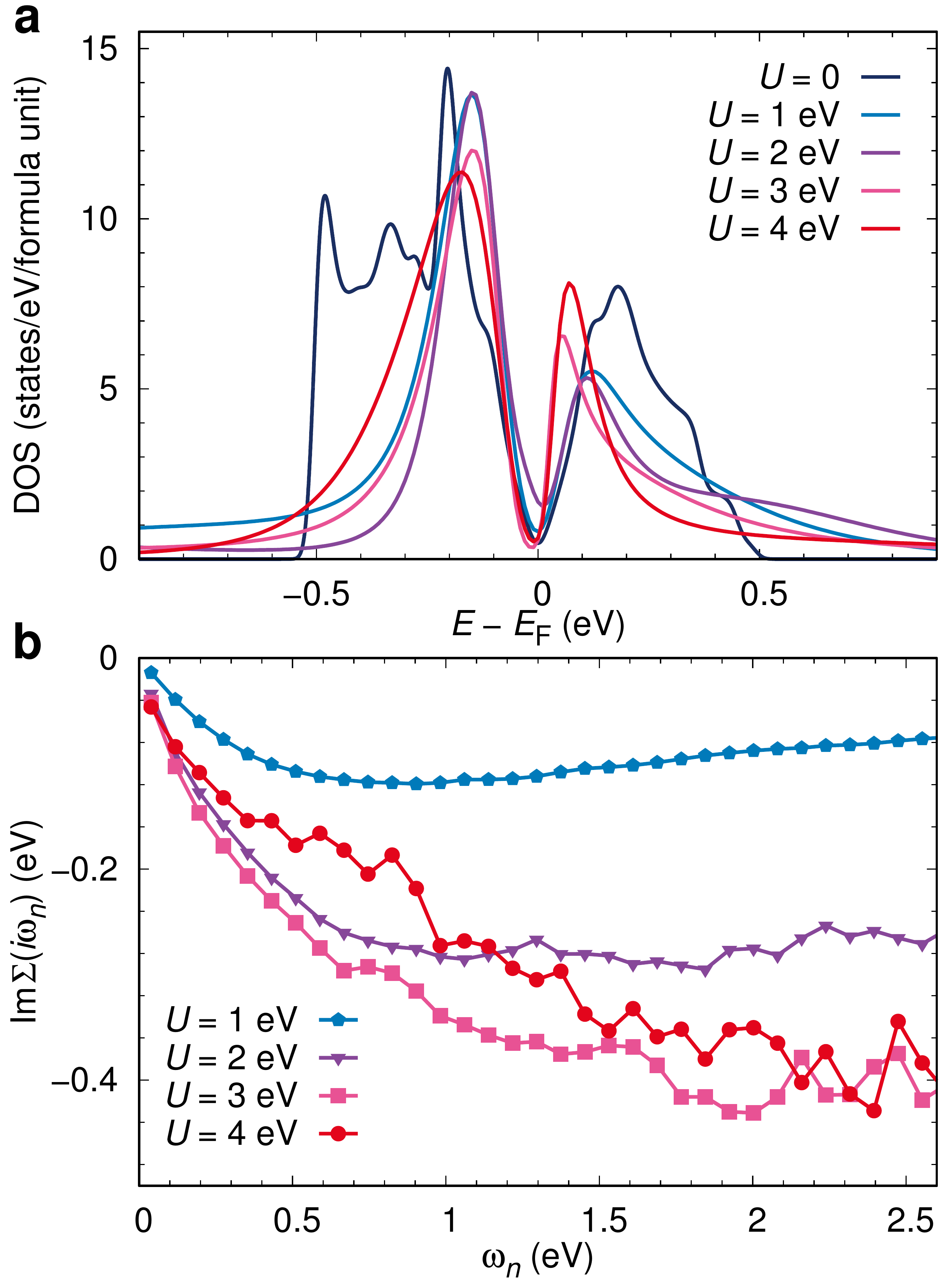}
\caption{\sffamily {\bf a} Density of states (DOS) at $T=1/80$~eV with
  finite doping $n=4/3$ in the paramagnetic regime. The DOS is
  obtained from 3-site DCA in combination with a CT-QMC impurity
  solver. The analytic continuation is performed using a maximum
  entropy method. {\bf b} Imaginary part of onsite self-energy as a
  function of Matsubara frequency at $T=1/80$~eV with the filling
  $n=4/3$ in the paramagnetic regime.} \label{fig:configuration}
\end{figure}

In order to account for quantum as well as short-range spatial
fluctuations beyond static mean-field theory, we employ the dynamical
cluster approximation (DCA) with $N_c=3$ sites in combination with an
interaction-expansion continuous-time quantum Monte Carlo approach as
an exact impurity solver~\cite{Rubtsov2005}.
Fig.~\ref{fig:configuration}~{\bf a} displays the density of states (DOS) for
$U=0$ to $U= 4$ eV for $T=1/80$ with finite doping of
$n=4/3$ in the paramagnetic regime.  We observe a band-narrowing
but the key low energy features, especially the absence of a gap with
very low DOS at $E_F$ reflecting the Dirac points, remain.

In order to see more clearly whether the pseudogap-like structure at
$E_F$ is related to the Dirac cone, we show the imaginary parts of the
self-energy as a function of Matsubara frequency for $U=1$~eV to
$U=4$~eV for $T=1/80$ with finite doping of $n=4/3$ in
Fig.~\ref{fig:configuration}~{\bf b}. The behaviour of the imaginary
part of the selfenergy follows Landau Fermi-liquid theory, where ${\rm
  Im}\,\Sigma(i\omega_n)$ approaches zero as $\omega_n\to 0$.  This
behaviour is also observed in DCA calculations for the honeycomb
lattice, where the Dirac cones are also
present~\cite{Liebsch2013}. This indicates that the onsite
interaction alone is insufficient to drive the system insulating at
the filling $n = 4/3$, in agreement with the results from the
DCA(RISB) approach. 

\subsection{DCA(RISB) details}

First, we applied DCA to the rotationally invariant slave-boson (RISB)
formalism~\cite{li89,lec07} in the saddle-point approximation. The
kagome lattice calls for the minimal cluster of three sites, which was
thus used for the calculations. Within the RISB scheme the electron's
quasiparticle character (fermionic $f_{\nu\sigma}$) and its
high-energy excitations (taken into account by the set of slave bosons
$\{\phi\}$) are decomposed on the operator level through
$\underline{c}_{\nu\sigma}$=$\hat{R}[\{\phi\}]_{\nu
  \nu^{\prime}}^{\sigma\sigma^{\prime}}f_{\nu^{\prime}\sigma^{\prime}}$,
where $\nu$ is a generic orbital/site index and
$\sigma$=$\uparrow,\downarrow$. The electronic self-energy
$\Sigma(\omega)$ at the saddle-point is local and incorporates terms
linear in frequency $\omega$, as well as static renormalisation
$\mathbf{\Sigma}^{\mathrm{stat}}$, i.e., $\mathbf{\Sigma
}(\omega)$=$(1-\mathbf{Z}^{-1})\omega+\mathbf{\Sigma}^{\mathrm{stat}}$,
whereby $\mathbf{Z}$ is the usual quasiparticle weight matrix. For
more details see Ref.~\onlinecite{lec07}. Note that the similarly
formulated single-site approach is very similar to the multi-orbital
Gutzwiller technique~\cite{bue98}.

\begin{figure*}[htb]
\includegraphics[width=0.85\textwidth]{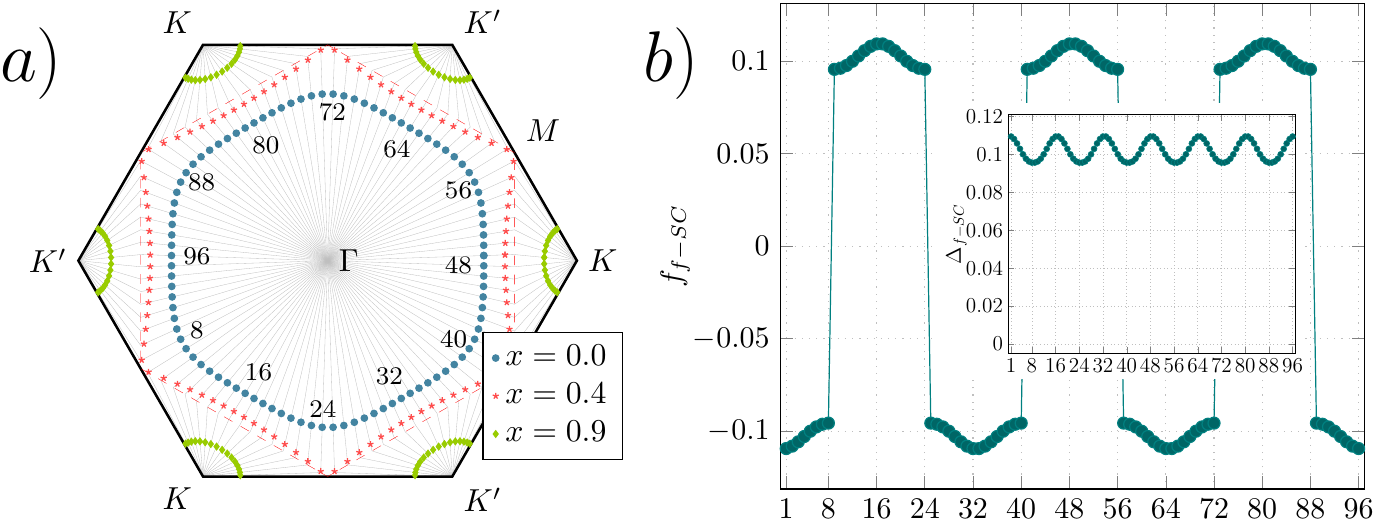}
\caption{\sffamily a) Model Fermi surfaces interpolating from Zn-type to Ga-type
filling. The Lifshitz transition happens from one contingent pocket centred
around $\Gamma$ to two pockets centred around $K$ and $K^{\prime}$. The
96-patch discretisation scheme for the FRG interaction vertex is
indicated. b) Characteristic scenario at electron-doped
Ga-herbertsmithite ($x=1$): $f$-wave is the dominant superconducting
instability with only small frustration in the Cooper channel. The
superconducting form factor is depicted along the patches of a). Due to the small pockets and the nodes located along
$\Gamma-M$, the SC gap looks rather homogeneous and changes sign as dictated
by the $B_{1u}$ irreducible lattice representation (inset: physical SC gap).}%
\label{fig:ga}%
\end{figure*}

\begin{figure}[htb]
\includegraphics[width=0.45\textwidth]{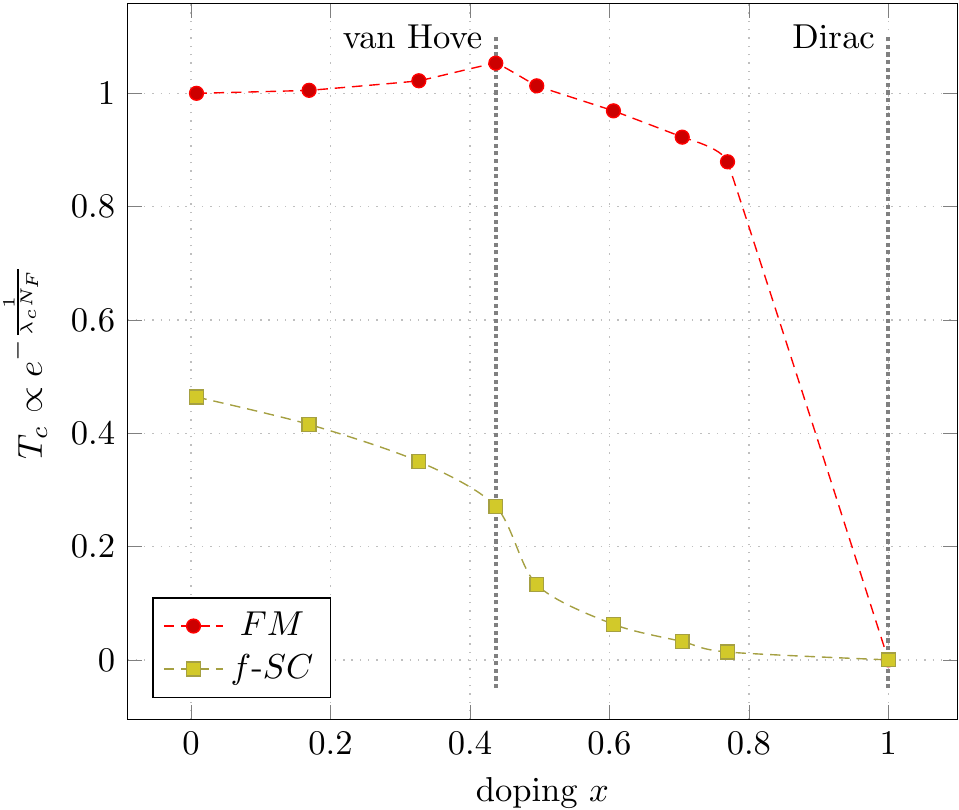}
\caption{\sffamily Critical scale $\Lambda_{c} \propto T_{c}$ for the
ferromagnetic fluctuations and $f$-wave superconducting instability (reference
scale chosen to be the ferromagnetic scale at half filling). The critical
scales are maximal around van Hove filling at the Lifshitz transition and
strongly decrease as the Fermi density of states decreases towards Dirac point
filling.}%
\label{fig:tc}%
\end{figure}

\subsection{FRG Details}

We employ multi-sublattice functional RG~\cite{pht} to investigate the
superconducting instability for a herbertsmithite Fermiology interpolating
from the Zn limit at half filling to the hole-doped Ga limit. We employ the
effective two-dimensional band structure obtained from our  ab initio
calculations.
Certainly, we cannot precisely identify at this stage for which range
$x<x_{c}$ the Mott state persists in the compound. Instead, for our RG
calculations, we assume a metallic normal state as a working hypothesis and
investigate the superconducting instabilities arising for different dopings.
Fig.~\ref{fig:ga}a shows the Fermiology assumed for different fillings from
$x=0$ to $x=1$. The functional RG necessitates the discretisation of the Fermi
surface into patches as depicted. There is a Lifshitz transition between a
single pocket centred around $\Gamma$ and two pockets centred around $K$ and
$K^{\prime}$, respectively, which shrink towards points as we approach the
Dirac point filling. There, the van Hove filling has been subject of intense
study as the enlarged density of states at the Fermi level might propagate
enhanced critical scales of unconventional Fermi surface
instabilities~\cite{PhysRevLett.110.126405,PhysRevB.87.115135}. However, as a
major difference to e.g. a similar Fermiology on the honeycomb lattice, it
should also be noted that the sublattice interference mechanism on the kagome
lattice~\cite{PhysRevB.86.121105} significantly contributes to suppressing
density wave instability with nested finite momentum transfer, providing a
further bias for ferromagnetic fluctuations to dominate at higher energies.


\FloatBarrier